\documentstyle[preprint,aps]{revtex}

\newcommand{\be}{\begin{equation}}
\newcommand{\ee}{\end{equation}}
\newcommand{\bea}{\begin{eqnarray}}
\newcommand{\eea}{\end{eqnarray}}
\newcommand{\ba}{\begin{array}}
\newcommand{\ea}{\end{array}}
\newcommand{\nn}{\nonumber \\}

\begin{document} 
\tightenlines

\preprint{}

\title{\vspace{2cm} \Large NONCOMMUTATIVE FIELD THEORIES AND GRAVITY }

\author{Victor O. Rivelles}
\address{Center for Theoretical Physics \\ Massachusetts Institute of
  Technology \\ Cambridge, MA  02139, USA \\ and \\
Instituto de F\'{\i}sica \\ Universidade de S\~{a}o Paulo\\
 Caixa Postal 66318, 05315-970, S\~{a}o Paulo, SP, Brazil\\
E-mail: rivelles@lns.mit.edu}

\maketitle
\thispagestyle{empty}

\begin{abstract}

We show that after the Seiberg-Witten map is performed the action for
noncommutative field theories can be regarded as a coupling to a field
dependent gravitational background. This gravitational background
depends only on the gauge field. Charged and uncharged fields couple to
different backgrounds and we find that uncharged fields couple more
strongly than the charged ones. We also show that the background is
that of a gravitational plane wave. A massless particle in this
background has a velocity which differs from the velocity of light
and we find that the deviation is larger in the uncharged case. This
shows that noncommutative field theories can be seen as ordinary
theories in a gravitational background produced by the gauge field
with a charge dependent gravitational coupling.

\end{abstract}

\newpage

Noncommutative (NC) theories have been studied in several contexts
since a long time ago. More recently it was found that they
arise as a limit of string theory with D-branes in a NS-NS background
B field \cite{Seiberg:1999vs}. In this limit gravity decouples but
still leaves some traces in the emerging NC field theory through the
Moyal product, defined as
\be
\label{Moyal}
A(x) \star B(x) = e^{\frac{i}{2} \theta^{\mu\nu} \partial^x_\mu
  \partial^y_\nu} A(x) B(y)|_{y\rightarrow x},
\ee
where $\theta^{\mu\nu}$ is the NC parameter. As a consequence, NC
theories are highly nonlocal and we would expect that they would be
troublesome. However, upon quantization, the ultraviolet structure is
not modified \cite{Filk:dm} but new infrared divergences appear and
get mixed with
the ultraviolet ones \cite{Minwalla:1999px}. This mixing of
divergences can be handled at
one loop level but when higher loops are taken into account the infrared
divergences are non-integrable turning the theory
nonrenormalizable. The only known exceptions for $d>2$ are supersymmetric
non-gauge theories \cite{Girotti:2000gc}. Even so this mixing of
divergences have
important consequences for many aspects of NC field theories
\cite{Szabo:2001kg}.
From a classical point of view many solutions from the commutative
field theory can be carried over to the NC corresponding
one. Instantons, monopoles and vortex solutions were found for the
NC Maxwell theory showing its resemblance with a non-Abelian
theory. The main feature of these solutions is that they are
non-singular and stable, properties usually not shared by their
commutative counterparts \cite{Harvey:2001yn}.

An important property of NC theories, which distinguishes them from
the conventional ones, is that translations in the NC
directions are equivalent to gauge transformations
\cite{Gross:2000ph}. This can be seen even for the case of a scalar
field\footnote{For the gauge field a translation is equivalent to a
  gauge transformation plus a constant shift of the potential
  \cite{Gross:2000ph}.}  which has the gauge transformation
$ \delta \hat{\phi} = -i [ \hat{\phi}, \hat{\lambda} ]_\star$,
where $[A,B]_\star  = A \star B - B \star A$ is the Moyal
commutator. Under a global
translation the scalar field transforms as $\delta_T \phi = \xi^\mu
\partial_\mu \hat{\phi}$. Derivatives of the field can be rewritten
using the Moyal commutator as $\partial_\mu \hat{\phi} = - i
\theta^{-1}_{\mu\nu} [ x^\nu, \hat{\phi}]_\star$ so that
$\delta \hat{\phi} = \delta_T \hat{\phi}$ with gauge parameter
$\hat{\lambda} = - \theta^{-1}_{\mu\nu} \xi ^\mu x^\nu$. The only other
field theory which has this same property is general relativity where
local translations are gauge transformations associated to general
coordinate transformations. This remarkable property shows that, as in
general relativity, there are no local gauge invariant observables in
NC theories.

An alternative approach to study NC theories makes use of commutative
fields (with its usual properties) instead of
the NC ones. They are related through the Seiberg-Witten (SW) map
\cite{Seiberg:1999vs} which is presented as a series expansion
in $\theta$. In this
way a local field theory is obtained at the expense of introducing
a large number of non-renormalizable interactions
\cite{Wulkenhaar:2002ps}. Quantization is 
problematic due to the number of divergences that appear. It seems
that at one loop level the SW map is just a field redefinition but at
higher loop orders this is not true \cite{Grimstrup:2002af}. At the
classical level, on the other 
side, it is possible to understand very clearly the breakdown of
Lorentz invariance induced by the noncommutativity. The dispersion
relation for plane waves in a magnetic background gets modified so
that photons do not move with the velocity of light
\cite{Guralnik:2001ax}.

We can wonder how other properties of NC field theories show up in the
commutative framework. In particular, the connection between
translations and gauge transformations seems to be lost. A global
translation on commutative fields can not be rewritten as a gauge
transformation. We will show in this paper that another aspect
concerning gravity emerges when commutative fields are
employed. Noncommutative field theories can be interpreted as ordinary
theories immersed in a gravitational background generated by the gauge
field. Firstly we notice that
the commutative theory can be regarded as an ordinary theory coupled
to a field dependent gravitational background. We will show that the
$\theta$ dependent terms in the commutative action can be interpreted
as a gravitational background which depends on the gauge field.
We then determine the metric which couples to real and
complex scalar fields. We find that the uncharged field coupling is
twice that of the charged one. So we can interpret the gauge coupling
in NC theory as a particular gravitational coupling which depends on
the charge of the field. We then show that the background describes a
gravitational plane wave. We also determine the geodesics
followed by  a massless particle in this background. We find that its
velocity differs from the velocity of light by an amount proportional
to $\theta$ with the deviation for the uncharged case being twice that
of the charged one. For the uncharged case the deviation is the same
as that found for the the gauge theory in flat space-time
\cite{Guralnik:2001ax,Cai:2001az}. As a final check we derive these
same velocities in a field theoretic context.

The action for the NC Abelian gauge theory in flat space-time is
\be
\label{NC_gauge_action}
S_A = -\frac{1}{4} \int d^4x \,\,\, \hat{F}^{\mu\nu} \star
\hat{F}_{\mu\nu},
\ee
where $\hat{F}_{\mu\nu} = \partial_\mu \hat{A}_\nu - \partial_\nu
\hat{A}_\mu - i [ \hat{A}_\mu, \hat{A}_\nu ]_\star$. For
a real scalar field in the adjoint representation of $U(1)$ the flat
space-time action is
\be
\label{NC_scalar_action}
S_\varphi = \frac{1}{2} \int d^4x \,\,\, \hat{D}^\mu \hat{\varphi} \star
\hat{D}_\mu \hat{\varphi},
\ee
where $\hat{D}_\mu \hat{\varphi} = \partial_\mu \hat{\varphi} - i
[\hat{A}_\mu, \hat{\varphi} ]_\star$.
On the other side, for a complex scalar field in the fundamental
representation of $U(1)$ the action is
\be
\label{NC_complex_scalar_action}
S_\phi = \int d^4 x \,\,\, \hat{D}^\mu \hat{\phi} \star ( \hat{D}_\mu
\hat{\phi} )^\dagger,
\ee
with $\hat{D}_\mu \hat{\phi} = \partial_\mu \hat{\phi} - i \hat{A}_\mu
\star \hat{\phi}$. The gauge transformations which leave the above
actions invariant are given by
\be
\label{gauge_transformations}
\delta \hat{A}_\mu = \hat{D}_\mu \hat{\lambda}, \qquad
\delta \hat{\varphi} = - i [ \hat{\varphi}, \hat{\lambda} ]_\star,
\qquad \delta \hat{\phi} = i
\hat{\lambda} \star \hat{\phi}, \qquad \delta \hat{\phi}^\dagger = -i
\hat{\phi}^\dagger \star \hat{\lambda}.
\ee

To go to the commutative framework we apply the SW map to the
fields. We assume that there exists a conventional Abelian gauge field
$A_\mu$ with the usual Abelian gauge transformation $\delta A_\mu =
\partial_\mu \Lambda$ such that $\hat{A}_\mu (A) +
\delta_{\hat{\Lambda}} \hat{A}_\mu (A) = \hat{A}_{\mu}(A+\delta_\Lambda
A)$. For the NC real scalar field $\hat{\varphi}$ we assume the
existence of a conventional uncharged scalar $\varphi$, with gauge
transformation $\delta \varphi = 0 $, such that $\hat{\varphi}
(\varphi,A) + \delta_{\hat{\Lambda}} \hat{\varphi} (\varphi,A) =
\hat{\varphi} (A+\delta_\Lambda A,\varphi+ \delta_\Lambda
\varphi)$. For the NC complex scalar field $\hat{\phi}$ we associate a
charged scalar field $\phi$ along the same lines. To first order in
$\theta$ we find
\bea
\label{SW_map}
\hat{A}_\mu &=& A_\mu - \frac{1}{2} \theta^{\alpha\beta}
A_\alpha ( \partial_\beta A_\mu + F_{\beta\mu} ), \nn
\hat{\varphi} &=& \varphi - \theta^{\alpha\beta} A_\alpha
\partial_\beta \varphi, \nn
\hat{\phi} &=& \phi - \frac{1}{2} \theta^{\alpha\beta} A_\alpha
\partial_\beta \phi,
\eea
We can now expand the NC actions
(\ref{NC_gauge_action}),(\ref{NC_scalar_action}) and
(\ref{NC_complex_scalar_action}) using (\ref{Moyal}) and apply the map
(\ref{SW_map}) to get the corresponding commutative actions.

For the real scalar field we find, always to first order in
$\theta$,
\be
\label{action_scalar_field}
S_\varphi = \frac{1}{2} \int d^4x \, \left[ \partial^\mu \varphi \partial_\mu
\varphi + 2 \theta^{\mu\alpha} {F_\alpha}^\nu \left( - \partial_\mu \varphi
\partial_\nu \varphi + \frac{1}{4} \eta_{\mu\nu} \partial^\rho \varphi
\partial_\rho \varphi \right) \right].
\ee
It is worth to remark that the tensor inside the parenthesis is
traceless. If we now consider this same field coupled to a
gravitational background
\be
\label{action_scalar_gravity}
S_{g,\varphi} = \frac{1}{2} \int d^4x \, \sqrt{-g} g^{\mu\nu}
\partial_\mu \varphi \partial_\nu \varphi,
\ee
and expand the metric $g_{\mu\nu}$ around the flat metric
$\eta_{\mu\nu}$,
\be
g_{\mu\nu} = \eta_{\mu\nu} + h_{\mu\nu} + \eta_{\mu\nu} h,
\ee
where $h_{\mu\nu}$ is traceless, we get
\be
\label{linearized_scalar_action}
S_{g,\varphi} = \frac{1}{2} \int d^4x \, \left( \partial^\mu \varphi
\partial_\mu \varphi - h^{\mu\nu}  \partial_\mu \varphi \partial_\nu
\varphi + h \partial^\rho \varphi \partial_\rho \varphi \right),
\ee
where indices are raised and lowered with the flat metric.
Since both actions, (\ref{action_scalar_field}) and
(\ref{linearized_scalar_action}), have the same structure we can
identify a linearized background gravitational field
\bea
\label{scalar_linearized_metric}
h^{\mu\nu} &= & \theta^{\mu\alpha} {F_\alpha}^\nu + \theta^{\nu \alpha}
{F_\alpha}^\mu + \frac{1}{2} \eta^{\mu\nu} \theta^{\alpha\beta}
F_{\alpha\beta},\nn
h &=& 0.
\eea
Then, the effect of noncommutativity on the commutative scalar field
is similar to a field dependent gravitational field.

The same procedure can be repeated for the complex scalar field. After
the SW map the action (\ref{NC_complex_scalar_action}) reduces to
\be
\label{SW-complex_scalar_action}
S_\phi =  \int d^4 x \, \left[ {D}^\mu {\phi} \,\, ( {D}_\mu
{\phi} )^\dagger - \frac{1}{2} \left( \theta^{\mu\alpha}
{F_\alpha}^\nu + \theta^{\nu\alpha} {F_\alpha}^\mu  + \frac{1}{2}
\eta^{\mu\nu} \theta^{\alpha\beta} F_{\alpha\beta} \right)
D_\mu \phi (D_\nu \phi)^\dagger \right].
\ee
Note again that the tensor inside the parenthesis is traceless. Now
the action for the charged scalar field in a linearized gravitational
field is
\be
\label{linearized_complex_scalar_action}
S_{g,\phi} = \int d^4 x \, \left[ {D}^\mu {\phi}  ( {D}_\mu
{\phi} )^\dagger - h^{\mu\nu} D_\mu \phi (D_\nu \phi)^\dagger + 2
h{D}^\mu {\phi}  ( {D}_\mu {\phi} )^\dagger \right],
\ee
which has the same structure as the action
(\ref{SW-complex_scalar_action}). Hence, the background gravitational
field in this case is
\bea
\label{linearized_metric_complex}
h^{\mu\nu} &=& \frac{1}{2} \left( \theta^{\mu\alpha} {F_\alpha}^\nu +
\theta^{\nu\alpha} {F_\alpha}^\mu \right) + \frac{1}{4} \eta^{\mu\nu}
\theta^{\alpha\beta} F_{\alpha\beta}, \nn
h &=& 0.
\eea
Then charged fields feel a gravitational background which is half of
that felt by the uncharged ones. Therefore, the gravity coupling is now
dependent on the charge of the field, being stronger for uncharged
fields. Notice that the gauge field has now a dual role, it couples
minimally to the charged field and also as a gravitational
background. 

We can now consider the gauge field. As it is well known the SW map
gives rise to the following action
\be
\label{SW_gauge_action}
S_A = -\frac{1}{4} \int d^4x \,\,\, \left[ F^{\mu\nu} F_{\mu\nu} + 2
\theta^{\mu\rho} {F_\rho}^\nu \left( {F_\mu}^\sigma F_{\sigma\nu} +
\frac{1}{4} \eta_{\mu\nu} F^{\alpha\beta} F_{\alpha\beta} \right)
\right].
\ee
Again, the tensor inside the parenthesis is traceless. At this point
we could be tempted to consider this action as some gravitational
action build up from the metric (\ref{scalar_linearized_metric}) or
(\ref{linearized_metric_complex}). Since the field strength always
appears multiplied by $\theta$ inside the metric, all invariants
constructed with it will be of order $\theta$. Hence, they can not
give rise to (\ref{SW_gauge_action}), unless they appear in
combinations involving the inverse of $\theta$. If we insist in having
an action which is polynomial in $\theta$ 
the best we can do is to regard the gauge field as having
a double role again and couple it to gravitation as in the previous
case. The linearized coupling of the Maxwell action is
\be
\label{maxwell_linear}
S_{g,A} = -\frac{1}{4} \int d^4x \,\,\, \left( F^{\mu\nu} F_{\mu\nu} +
  2 h^{\mu\nu} {F_\mu}^\rho F_{\rho\nu}  \right).
\ee
Since it does not couple to the trace part of the metric $h$ remains
arbitrary and $h^{\mu\nu}$ is given by
  (\ref{linearized_metric_complex}). Since the NC gauge field
resembles a non-Abelian gauge field we expect that its commutative
counterpart couple to the same gravitational field as the charged
one. It should also be remarked that in this case the
gravitational field can not be interpreted just as a fixed
background since it depends on the dynamical gauge field.


Having determined the field dependent background metric we can now
study its properties. We will consider the metric which couples to the
charged fields (\ref{linearized_metric_complex}). To consider the
metric (\ref{scalar_linearized_metric}) we have just to replace
$\theta$ by $2\theta$. The linearized metric is then
\be
\label{metric}
g_{\mu\nu} = \eta_{\mu\nu} + \frac{1}{2} \left( \theta_{\mu\alpha}
{F^\alpha}_\nu + \theta_{\nu\alpha} {F^\alpha}_\mu \right) +
\frac{1}{4} \eta_{\mu\nu} \theta^{\alpha\beta} F_{\alpha\beta}.
\ee
The geometric quantities can be evaluated without difficulty and we find
\bea
\label{riemann}
R_{\mu\nu\rho\sigma} &=& \frac{1}{2} [- \theta_{\alpha[\mu}
  \partial_{\nu]} \partial^\alpha F_{\sigma\rho} + \theta_{\rho\alpha}
  \partial^\alpha \partial_{[\mu} F_{\nu]\sigma} +
  \theta_{\sigma\alpha} \partial_\rho \partial_{[\mu}
  {F^\alpha}_{\nu]} \nn
 &+& \theta^{\alpha\beta} \left( \eta_{\sigma[\mu}
  \partial_{\nu]} \partial_\rho F_{\alpha\beta} - \eta_{\rho[\mu}
  \partial_{\nu]} \partial_\sigma F_{\alpha\beta} \right) ],
\eea
\be
\label{ricci}
R_{\mu\nu} = \frac{1}{4} \left( {\theta_\mu}^\alpha \partial_\alpha
\partial^\beta F_{\beta\nu} + {\theta_\nu}^\alpha \partial_\alpha
\partial^\beta F_{\beta\mu} + \frac{1}{2} \eta_{\mu\nu}
\theta^{\alpha\beta} \Box F_{\alpha\beta} \right),
\ee
\be
\label{scalar_curvature}
R = \frac{1}{4} \theta^{\alpha\beta} \Box F_{\alpha\beta}.
\ee
Notice that all of them, and also the Christoffel symbol, are first
order in $\theta$. Since on-shell and in the absence of matter
$\partial^\mu F_{\mu\nu}$ is first order in $\theta$, then $\Box
F_{\mu\nu}$ is also first order\footnote{Notice that field equations
  for the gauge field are derived from (\ref{SW_gauge_action}) which
  is defined in flat space-time. Hence in the field equations and
  solutions the Minkowski metric is used.}. This means that the Ricci
tensor and the scalar curvature both vanish but not the Riemann tensor
so that the metric (\ref{metric}) is not that of a flat
space-time. Then, to order zero in $\theta$,
$F_{\mu\nu}$ satisfies the wave equation and is a function of $k_\mu
x^\mu$ with $k^2=0$. Hence, the metric has all symmetries of a
gravitational plane wave.

More rigorously, we find that the noncommutative parameter is
covariantly conserved $D_\mu \theta^{\alpha\beta} = 0$. We then have a
geometry equipped with a covariantly constant two-form. Since
$\theta^{\alpha\beta} \theta_{\alpha\beta} = 0$ to first order then
the two-form is also null. The existence of this covariantly null two-form
guarantees that the metric (\ref{metric}) describes a pp-wave
\cite{Ehlers}. More than that, if we consider the solution for the
gauge field to first order in $\theta$ and in the absence of matter as
$F_{\mu\nu} = k_{[\mu} F_{\nu]}$, with $k^\mu$ a null vector and
$F_\mu$ transversal, $k^\mu F_\mu = 0$, then
$\partial_\alpha R_{\mu\nu\rho\sigma} = k_\alpha R_{\mu\nu\rho\sigma}$
and the complex Riemann tensor
\be
\label{comples_riemann}
P_{\mu\nu\rho\sigma} = R_{\mu\nu\rho\sigma} + \frac{i}{2}
{\epsilon_{\rho\sigma}}^{\alpha\beta} R_{\mu\nu\alpha\beta}
\ee
satisfies $\partial_\alpha P_{\mu\nu\rho\sigma} = k_\alpha
P_{\mu\nu\rho\sigma}$. This shows that the pp-wave is in fact a
plane wave \cite{Ehlers}. Then the metric (\ref{metric}) is that of a
gravitational plane wave.

We can now turn our attention to the behavior of a massless particle
in this background. Its geodesics is described by
\be
\label{ds}
ds^2 = \left( 1 + \frac{1}{4} \theta^{\alpha\beta} F_{\alpha\beta}
\right) dx^\mu dx_\mu + \theta_{\mu\alpha} {F^\alpha}_\nu dx^\mu
dx^\nu = 0.
\ee
If we consider the case where there is no noncommutativity between
space and time, that is $\theta^{0i}=0$, and
calling $\theta^{ij} = \epsilon^{ijk} \theta^k$, $F^{i0} = E^i$, and
$F^{ij} = \epsilon^{ijk} B^k$, we find to first order in $\theta$ that
\be
\label{photon_eq}
( 1 -  \vec{v}^2 )( 1 - 2 \vec{\theta} \cdot \vec{B}) - \vec{\theta}
\cdot (\vec{v} \times \vec{E} ) + \vec{v}^2 \vec{\theta} \cdot \vec{B}
- (\vec{B} \cdot \vec{v}) ( \vec{\theta} \cdot \vec{v}) = 0,
\ee
where $\vec{v}$ is the particle velocity. Then to zeroth order, the
velocity $\vec{v}_0$ satisfies
$\vec{v}_0^2 = 1$ as it should. We can now decompose all vectors into
their transversal and longitudinal components with respect to
$\vec{v}_0$, $\vec{E} = \vec{E}_T + \vec{v}_0 E_L$, $\vec{B} =
\vec{B}_T + \vec{v}_0 B_L$ and $\vec{\theta} = \vec{\theta}_T + \vec{v}_0
\theta_L$. We then find that the velocity is
\be
\label{velocity_complex_scalar}
\vec{v}^2 = 1 + \vec{\theta}_T \cdot ( \vec{B}_T - \vec{v}_0 \times
\vec{E}_T ).
\ee
Hence, a charged massless particle has its velocity changed with
respect to the velocity of light by an amount which depends on
$\theta$. For an uncharged massless particle
\be
\label{velocity_real_scalar}
\vec{v}^2 = 1 + 2 \vec{\theta}_T \cdot ( \vec{B}_T - \vec{v}_0 \times
\vec{E}_T ),
\ee
and the correction due to the noncommutativity is twice that of a charged
particle.

We can now check the consistency of these results by going back to the
original actions (\ref{action_scalar_field}) and
(\ref{SW-complex_scalar_action}), and
computing the group velocity for planes waves. Upon quantization
they give the velocity of the particle associated to the
respective field. For the uncharged scalar field we get the equation
of motion
\be
\label{eq_motion_real_scalar}
\left( 1 - \frac{1}{2} \theta^{\mu\nu} F_{\mu\nu} \right) \Box
\varphi - 2 \theta^{\mu\alpha} {F_\alpha}^\nu \partial_\mu\partial_\nu
\varphi = 0.
\ee
If the field strength is constant we can find a plane wave solution
with the following dispersion relation
\be
\label{dispersion_relation_real_scalar}
\left(  1 - \frac{1}{2} \theta^{\mu\nu} F_{\mu\nu} \right) k^2 - 2
\theta^{\mu\alpha} {F_\alpha}^\nu k_\mu k_\nu = 0,
\ee
and using the same conventions for vectors as before, it results in
\be
\label{dispersion_relation_vector_form}
\frac{\vec{k}^2}{\omega^2} = 1 - 2 \vec{\theta}_T \cdot ( \vec{B}_T -
\frac{\vec{k}}{\omega} \times \vec{E}_T ),
\ee
where $k^\mu = (\omega, \vec{k})$.
We then find that the phase and group velocities coincide and are given
by (\ref{velocity_real_scalar}) as expected. For the charged scalar
field we have to turn off the gauge coupling in order to get a plane
wave solution. In this case the equation of motion is
\be
\label{eq_motion_complex}
\left( 1 - \frac{1}{4} \theta^{\mu\nu} F_{\mu\nu} \right) \Box
\phi - \theta^{\mu\alpha} {F_\alpha}^\nu \partial_\mu\partial_\nu
\phi = 0.
\ee
In a constant field strength background the dispersion relation for a
plane wave reads as in (\ref{dispersion_relation_real_scalar}) with
$\theta$ replaced by $\theta/2$. Then we must perform the same
replacement in the phase and group velocities and we get
(\ref{velocity_complex_scalar}). Therefore, in both pictures,
noncommutative and gravitational, we get the same results.

For the gauge field the situation is more subtle because of its double
role. There is no clear way to
split the action (\ref{SW_gauge_action}). What can be done is to
break up the gauge field into a background plus a plane wave as 
in \cite{Guralnik:2001ax}. We then get the following dispersion
relation
\be
\label{dispersion_relation_gauge}
k^2 - 2 \theta^{\mu\alpha} {F_\alpha}^\nu k_\mu k_\nu = 0,
\ee
where ${F_\alpha}^\nu$ is now the constant background. This
leads to (\ref{dispersion_relation_vector_form}), that is, the
dispersion relation for the uncharged scalar field. It also reproduces
the result in \cite{Guralnik:2001ax,Cai:2001az} when the background is purely
magnetic. This shows the dual role of the gauge field, since it
couples to gravitation as a charged field but its dispersion relation
is that of an uncharged field. 

We have seen in this paper that it is possible to regard
noncommutative theories as conventional theories embedded in a
gravitational background produced by the gauge field. This brings a
new connection between noncommutativity and gravitation. We could
imagine that this is a peculiarity of the first order term in the
$\theta$ expansion of the SW map. If we consider the SW map for the
scalar field to second order in $\theta$ we find that it is linear in
the scalar field. A short calculation shows that the action
(\ref{NC_scalar_action}) remains quadratic after the SW map and that
it can be cast again in the form (\ref{action_scalar_gravity}) since
all terms of the form  $\partial^2 \varphi \partial^2 \varphi$
cancel. Then it seems to be possible to keep the same interpretation
to all orders in $\theta$. 


I would like to thank R. Jackiw for discussions and comments on the
manuscript. This work was partially supported by CAPES and PRONEX
under contract CNPq 66.2002/1998-99.

\samepage

\end{document}